\documentclass[10pt]{article}

\usepackage{amsmath}
\usepackage{amssymb}

\usepackage{nameref,hyperref}

\usepackage{graphicx}

\usepackage{cite}

\usepackage{color} 

\usepackage{fixltx2e}

\usepackage{color}

\usepackage[labelfont=bf,labelsep=period,justification=raggedright]{caption}

\makeatletter
\renewcommand{\@biblabel}[1]{\quad#1.}
\makeatother

\date{}

\begin{document}

\begin{flushleft}
{\Large
\textbf\newline{Imitation Combined with A Characteristic Stimulus Duration Results in Robust Collective Decision-making}
}
\newline
\\
Sylvain Toulet\textsuperscript{1,2},
Jacques Gautrais\textsuperscript{1,2},
Richard Bon\textsuperscript{1,2},
Fernando Peruani\textsuperscript{3,*}\footnote{Email: peruani@unice.fr}
\\
\bigskip
\bf{1} Universit{\'e} de Toulouse, UPS, Centre de Recherches sur la Cognition Animale, 118 route de Narbonne, F-31062 Toulouse Cedex 9, France.
\\
\bf{2} CNRS, Centre de Recherches sur la Cognition Animale, 118 route de Narbonne, F-31062 Toulouse Cedex 9, France.
\\
\bf{3} Laboratoire J. A. Dieudonn{\'e}, Universit{\'e} de Nice Sophia Antipolis, UMR 7351 CNRS, Parc Valrose, F-06108 Nice Cedex 02, France.
\\
\bigskip

%




\end{flushleft}
\section*{Abstract}
For group-living animals, reaching consensus to stay cohesive is crucial for their fitness, particularly when collective motion starts and stops. Understanding the decision-making at individual and collective levels upon sudden disturbances is central in the study of collective animal behavior, and concerns the broader question of how information is distributed and evaluated in groups. Despite the relevance of the problem, well-controlled experimental studies that quantify the collective response of groups facing disruptive events are lacking. Here we study the behavior of groups of uninformed individuals subject to the departure and stop of a trained conspecific \textcolor{black}{within small-sized groups}. We find that the groups reach an effective consensus: either all uninformed individuals follow the trained one (collective motion occurs) or none does. Combining experiments and a simple mathematical model we show that the observed phenomena results from the interplay between simple mimetic rules and the characteristic duration of the stimulus, here, the time the trained individual is moving away. The proposed mechanism strongly depends on group size, as observed in the experiments, and though group splitting can occur, the most likely outcome is always a coherent collective group response (consensus). The prevalence of a consensus is expected even if the groups of naives face conflicting information, e.g. if groups contain two subgroups of trained individuals, one trained to stay and one trained to leave. Our results indicate that collective decision-making and consensus in (small) animal groups are likely to be self-organized phenomena that do not involve concertation or even communication among the group members.


\section*{Introduction}
\textcolor{black}{Animals on the move show an impressive capacity to respond to strong perturbation such as changes of directions or behavioral switches \cite{Attanasi:2015tm,Ioannou:2015dea}}. Many gregarious vertebrates are fusion-fission species, with frequent changes in size and composition of groups. In addition, the habitat they live in is generally heterogeneous, such that individuals alone or in groups may have to move among more or less distant areas in order to fulfil their basic vital requirements, e.g. drinking, resting or avoiding pblackators \cite{Beauchamp:2012hm,Conradt:2005di,Fischhoff:2007is,King:2009uz,Sueur:2011io}. Even during feeding periods, animals have to search for available food, moving slowly and on short distances between feeding stations or more rapidly and on larger distances to exploit distinct feeding areas \cite{Dumont:2000tc}. Thus, groups either resting or feeding with animals motionless or moving slowly are frequently joined by incoming individuals but also submitted to departures of group members \cite{Reebs:2000ta,Ramseyer:2009bk,Ward:2013gz}.
When faced to multiple choices, social organisms must reach consensus in order to maintain the cohesion of the group and the advantages linked to it \cite{Fischhoff:2007is,Danchin:2004kz,GalefJr:2001kl}. 
The departure of one or few individuals from static groups as well as stops in moving groups compromise social cohesion \cite{Couzin:2002em,Eftimie:2007gs}. 
This is particularly critical for small groups -- a scenario that applies to most gregarious animals \cite{Reiczigel:2008fua} 
despite the popularity and fascination that produce giant bird flocks or fish schools \cite{Ballerini:2008tv} --  
where group splitting represents a serious pblackatory risk \cite{Beauchamp:2012hm}. 

What influences the individual decisions, \textit{i.e} the interplay between external stimuli and internal state, and which decision-making processes occur to maintain social cohesion are among the most compelling questions in the study of collective animal behavior \cite{Sueur:2011io,Conradt:2010er,Giardina:2008ff,King:2010gt,Petit:2010ko,Procaccini:2011cb,Roberts:1997ie,Ward:2008cx,Collignon:2012dz, Lihoreau:2010ka}. \textcolor{black}{This is particularly true in very large groups on the move like flocks of birds where the propagation of information through local interactions plays a key role in the form of the collective response \cite{Attanasi:2014fc,Cavagna:2014vn}. This emerges from the very large number of individuals or the strong effects of density in the group that implies that localized responses must be implied to reach collective cohesive decisions.} This goes beyond biology and concerns the broader question of how information at the individual level is evaluated, processed and distributed in the group \cite{Bellman:1970dp,Dyer:2008dh,Edwards:1954eq,Herrera:2005ef,Lindley:1985tw}. Recent experiments with primates and fish revealed that an individual spontaneously departing from a static group is likely to give up and return to the group when not followed \cite{Reebs:2000ta,Roberts:1997ie}. It has been also observed that collective motion is promoted by already moving conspecifics \cite{Ward:2008cx,Petit:2009bz}. Finally, in some species, a quorum is requiblack to observe collective movement \cite{Ward:2008cx,Petit:2009bz,Pratt:2005bp,Sumpter:2008bw,Sempo:2009cp}. On the other hand, how collective motion stops, remains largely unexploblack except for few exceptions \cite{Petit:2010ko,Bhattacharya:2010kh,Daruka:2009jr}. In summary, there is a lack of experimental and theoretical studies focusing on the decision-making processes that trigger and stop collective motion \cite{Beauchamp:2012hm,Conradt:2005di,King:2009uz,Cresswell:2000iq}. 

Here, we use experimentally acquiblack data to address how groups of uninformed individuals respond to the departure and stop of an informed conspecific \textcolor{black}{in small-sized groups where we suppose that individuals have a global perception of all the group members and that effects of group density are negligible}. We show that the groups of uninformed individuals always reach a consensus: either all uninformed individuals follow the informed one or none does it. 
Combining experiments where we control the stimulus, associated to the motion of the informed individual, and a mathematical model we unveil that the apparent collective decision-making process leading to an effective consensus results from a self-organized phenomenon resulting from the interplay of simple mimetic rules and the characteristic duration of the stimulus, with group size playing a central role. 

\section*{Materials and Methods}

\subsection*{Study area and Experiments}
	Fieldwork was carried out in the Domaine du Merle ($5.74^\circ$ E , $48.50^\circ$ N) in the south of France. The field station is located in the Crau region, a very flat area coveblack by a native steppe. The experiments were performed within irrigated pastures providing homogeneous food resources. From the available flock of 1400 ewes, 30 of them were randomly selected and allocated to the training set and a further 200 ewes to the naive set. All ewes used were unrelated and were 3 to 5 year-old. A number was painted on the back and fleece of each sheep for identification. Both sets were penned up in the same sheepfold during the evening and the night. All the experiments were carried out in daylight (from 10am to 6pm) and the ewes were fed hay in the sheepfold in the morning and in the evening.

\subsection*{Sheep training}
	The experiments rely on our capacity to trigger the movement of one sheep toward a fixed panel at the periphery of an arena at a desiblack time. The protocol used to trained sheep was similar to the one used in previous experiments \cite{Pillot:2010dx,Pillot:2011jz}. Sheep to be trained were originally allocated at random in 6 groups of 5 animals which composition remained unchanged during the training period. Sheep were first habituated in the sheepfold to feed on corn and to receive simultaneously a vibration provided by a neck collar during 3 days. Then the training groups were introduced successively in one of two test arenas (50 x 50m), for a period of 30 to 40 min, each animal wearing a vibrating collar. Ten minutes past the introduction, the collar was activated and one yellow panel (0.5 x 0.5m) was simultaneously raised delivering a handful of corn. Each group received four to eight stimulations, each separated by a period of at least 5 min during each training session. Past 14 days of training, we selected the 3 sheep with the best learning scores (100\% of departure toward the panel following a vibration). \textcolor{black}{These 3 trained sheep were comparable in terms of initiation behavior and did not show any differences in movement speed to the panel (ANOVA: $F^{^{2}}_{_{45}} = 0.378, P = 0.68$).}
	Meanwhile, the set of naive sheep to be used in experiments was confronted to panel rising (without corn delivery) at the periphery of experimental arenas, at one-min interval during two sessions of 90 min. At the end of this habituation session, no naive sheep raised its head when raising the panel. In addition, these two days allowed naive sheep to be familiarized with the experimental setup.
	
\subsection*{Experimental procedure}
	The experimental setup consisted in two arenas (50 x 50m) delimited with sheep fences and surrounded by a visual barrier (propylene net). A 7m-high tower was placed at an equal distance (10m) apart from two next arenas’ corners. Yellow panels (0.5 x 0.5 m) were hammeblack in the middle of each side for both arenas and were not visible to sheep (\nameref{S1_Fig}). The tests consisted in introducing groups of 8, 16 or 32 sheep within one arena, among which one trained sheep equipped with a vibrating collar. The trained ewes were used no more than twice a day, but were implicated in all group sizes. The naive sheep that composed the rest of the groups were selected randomly for each test. Because of the large number of individuals needed to complete all replications, the naive sheep were used several times, except in groups of 8. Thirty replications were performed for groups of 32 and 15 replications for groups of 8 and 16 individuals. A test was conducted as follows: the group was introduced in the arena and sheep grazed spontaneously during 20 min. Then, one of the two panels closest to the tower was raised, waiting for all sheep grazing (\textit{i.e} head down). Simultaneously the vibrating collar of the trained sheep was activated for 2 sec. Past 10 min (end of test), a new panel was raised (one of the two farthest from the tower) to reinforce the conditioning of trained ewes and avoid restricting their space use to the vicinity of panels closest to the tower. The group was led back to the sheepfold shortly afterward. The naive sheep that were not tested during one experimental day were introduced in distant pasture. We never performed two trials in parallel.
		We also carried out control experiments to be sure that naive ewes did not associate the panel rise and the food reward. Thereby, 6 tests before and 6 after the test series were conducted with groups of 32 naive ewes, using the same protocol as described before. We found no movement of groups when raising the panel, almost all sheep continuing their spontaneous activity.
\subsection*{Data collection and analyses}
Two digital cameras (Canon EOS D50) were fixed on the tower, each one focusing on one arena. Fifteen minutes after the introduction of the groups, the digital camera was turned on, taking a picture of the entire arena every second and turned off five minutes after the panel was raised. For each replication, we obtained a series of about 600 pictures. Using a custom software developed by JG \textcolor{black}{\cite{Ginelli:fc,Michelena:2008ej}}, we were able to track on each picture the position and the orientation of animals by dragging a vector on their back, and identify the behavior of each individual, \textit{i.e} grazing, standing head-up, moving and others.

We defined a departure of the trained sheep (initiator), \textit{i.e} initiation past experimental stimulation when it performed an uninterrupted walk towards the raised panel. The following behavior, \textit{i.e} a new departure, was defined as the movement of a naive ewe occurring after the trained sheep departure, without stop until joining the trained ewe near the panel. The behavior of stopping was defined as an individual ceasing to walk and remains either stationary head-up or resumes grazing. Six replication in the groups of 32 were discarded, one because the initiator did not depart, two because the initiator stopped moving between the group and the target and three because the initiator showed a moving behavior not comparable to other trials (going to a wrong target first and then joining the rewarded target). Thus we performed analyses on 15 trials for groups of 8 and 16 sheep and 24 trials for groups of 32.

The level $\alpha$ was fixed to 0.05 for statistical significance. All analyses were conducted using R version 3.0.1.

\subsection*{Ethics statement}
All the animals were maintained under routine husbandry conditions at a Montpellier Supagro research station (Domaine du merle, Salon-de-Provence, France) with full approval of its director Pierre-Marie Bouquet. Animal welfare requirements were fully respected in accordance with the European Directive 2010/63/EU, with the rules of the European Convention for the Protection of Vertebrate Animals used for Experimental and Other Scientific Purposes and with the Convention of the French Comit\'e national de r\'eflexion \'ethique sur l’exp\'erimentation animale. No special authorization from the French Ethical Committee for animal experimentation (Commission nationale de l'expérimentation animale) was requiblack as no protected or endangeblack species was involved, as the experiments did not imply any invasive manipulation (the experimental protocol consists in the observation of groups and the acquiblack data are only pictures of the animals) and as sheep were conducted to the test arenas, as they are herded on a daily basis to the pastures. All personnel involved had technical support from the employees of the research station as requiblack by the French Ministry of Research. The experimental protocols included short test periods (20 minutes) where sheep did not experience painful, stressful or unfamiliar situations. The experimental procedures had no detrimental effect on the sheep and at the end of the experiment all the animals reintegrated the sheep herd of the breeding research station.

\section*{Results and Discussion}
\begin{figure}[h]
\centering\resizebox{11 cm}{!}{\rotatebox{0}{\includegraphics{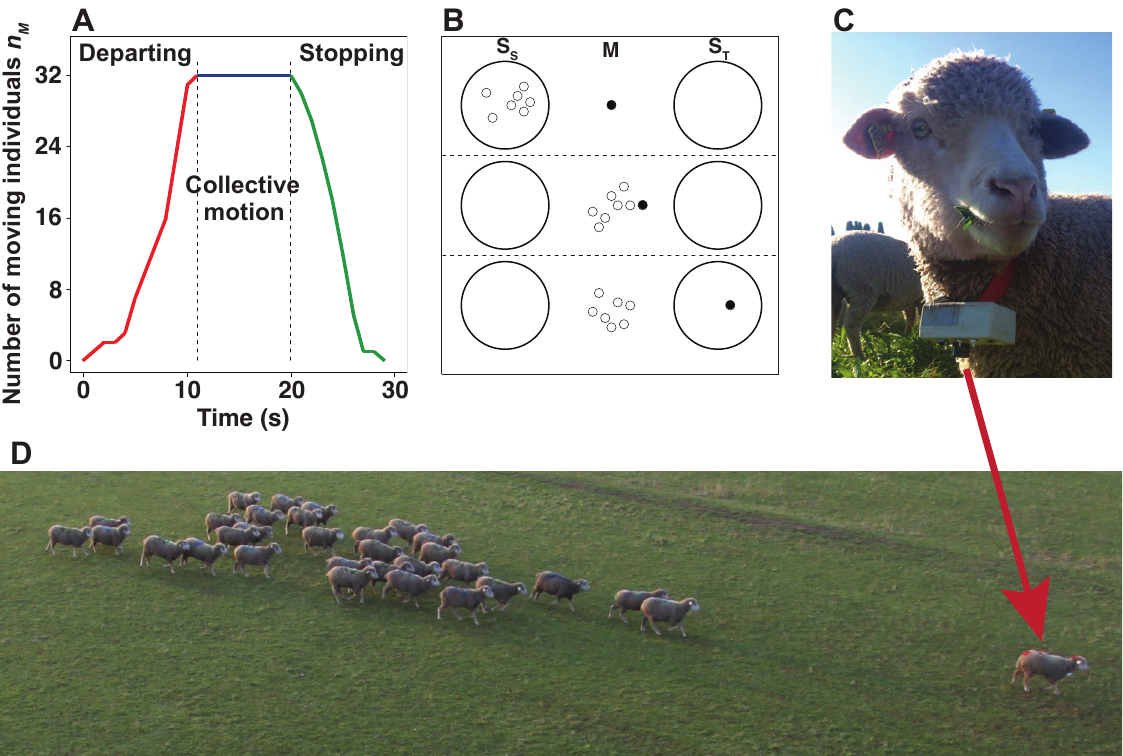}}}
\caption{{\bf Experimental collective observations.}
(A) The number of moving individuals (n\textsubscript{M}) as a function of time in one of the trials with 32 sheep. The departing (in black), collective motion (in blue) and stopping (in green) phases are indicated. (B) Sketch illustrating the temporal phases of an experiment. The three behavioral states of individuals are represented: (static) individuals at the starting position (S\textsubscript{S}), moving (M), and stopped near the target (S\textsubscript{T}).  The initiator is depicted by a full circle, while open circles correspond to naive individuals. From top to bottom, we observe the first transition S\textsubscript{S} $\rightarrow$ M, the collective motion phase, and the first transition  M $\rightarrow$ S\textsubscript{T}. (C) shows one of the trained individual fitted with the vibrating collar. 
(D) A snapshot of a herd of 32 sheep in collective motion provoked by the departure of the trained individual.}
\label{fig1}
\end{figure}

In our experiments, we work with groups of N~=~8, 16 and 32 sheep, among which 1 is a trained individual -- henceforth referblack to as initiator -- while the remaining N - 1 are uninformed/naive individuals. 
The initiator is trained to move towards a target located at the periphery of the arena when a vibrating collar is activated by a remote control (Fig.~\ref{fig1}C). The group is subject  to the perturbation produced 
by the initiator: {\it i.e.} the sudden departure and stop of the initiator, which challenge the social cohesion of the group. When the departure of the initiator triggers a collective response, 
we observe  three distinct consecutive phases:  departing, collective motion, and stopping as illustrated in Fig.~\ref{fig1}A.
The departing phase starts with the departure of the initiator and continues until the number n\textsubscript{M} of moving individuals increases to match the group size. At this point, the collective motion phase begins with the group moving cohesively behind the initiator. The stop of the initiator near the target marks the onset of the stopping phase, where n\textsubscript{M} decreases until reaching 0. The behavior of an individual can be characterized by one of the following states: stopped at the start position (S\textsubscript{S}), moving (M), or stopped at the target position (S\textsubscript{T}) (Fig.~\ref{fig1}B). The whole process can be then described as a transition first from S\textsubscript{S} to M, and then from M to S\textsubscript{T}. 

In all trials with 8 and 16 sheep, the departure of the initiator systematically triggers a collective motion (Fig.~\ref{fig1}D). In groups of 32 
the departure of the initiator does not always leads to a collective motion of the naive group. Splitting of the naive group has not been observed (see \nameref{S1_Video} and \nameref{S2_Video} for examples.)  

\begin{figure}[h]
\centering\resizebox{11 cm}{!}{\rotatebox{0}{\includegraphics{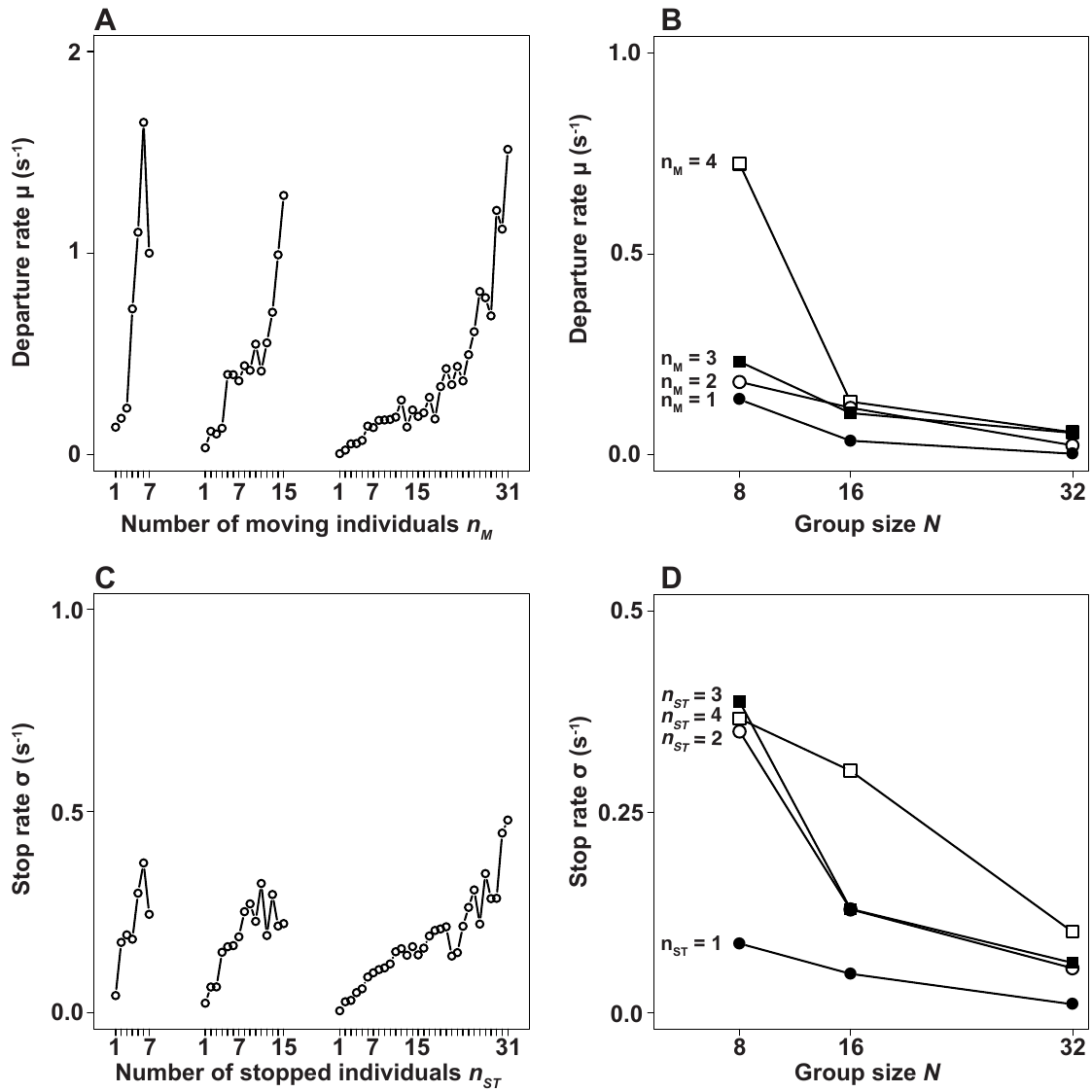}}}
\caption{{\bf Individual transition rates.}
(A) and (C) show  that the departure and stop rate increase with n\textsubscript{M} and n\textsubscript{S\textsubscript{T}}, respectively, for all group sizes (N~=~8, 16 and 32). (B) The inhibiting effect of S\textsubscript{S} on the transition S\textsubscript{S} $\rightarrow$ M for a given n\textsubscript{M} is evidenced by the decrease of the transition rate with N. (D) Similarly, in the transition M $\rightarrow$ S\textsubscript{T}, we observe a decrease of the transition rate with N for a fixed value of n\textsubscript{S\textsubscript{T}}, which indicates an inhibiting role of n\textsubscript{M}.}
\label{fig2}
\end{figure}

We start our analysis by focusing first on  the cases where the initiator successfully provokes a collective motion. We are interested in quantifying the decision of sheep to switch from S\textsubscript{S} to M and from M to S\textsubscript{T}. 
\textcolor{black}{To account for the dynamics of the departing and stopping phases, we focused on the individual transition rates (the probability per unit of time for a given individual to switch behavior). From the experimental data we estimated the departure and stopping rates for each departure and stopping rank (see \nameref{SI_mu_Text} for details of the computation).}
Fig.~\ref{fig2}A (respectively, Fig.~\ref{fig2}C) shows that \textcolor{black}{the transition rate} from from S\textsubscript{S} to M (in Fig.~\ref{fig2}C, from M to S\textsubscript{T}) increases with n\textsubscript{M} (with n\textsubscript{S\textsubscript{T}}, the number of individuals in state S\textsubscript{T}, in Fig.~\ref{fig2}C). Fig.~\ref{fig2}B (respectively, Fig.~\ref{fig2}D) indicates that the transition rate from S\textsubscript{S} to M (respectively, from M to S\textsubscript{T}) for a fixed value of n\textsubscript{M} (fixed value n\textsubscript{S\textsubscript{T}}, for M $\rightarrow$  S\textsubscript{T}) diminishes with n\textsubscript{S\textsubscript{S}} (n\textsubscript{M} in Fig.~\ref{fig2}D). 
This indicates that both transitions S\textsubscript{S} $\rightarrow$ M and M $\rightarrow$ S\textsubscript{T} share similar features: they both exhibit a promoting component (n\textsubscript{M} in S\textsubscript{S} $\rightarrow$ M and n\textsubscript{S\textsubscript{T}} in M $\rightarrow$ S\textsubscript{T}) and an inhibiting component (n\textsubscript{S\textsubscript{S}} in S\textsubscript{S} $\rightarrow$ M and n\textsubscript{M} in M $\rightarrow$ S\textsubscript{T}). This is consistent with previous studies using smaller group sizes \cite{Pillot:2010dx,Pillot:2011jz}.

\begin{figure}[h]
\centering\resizebox{11 cm}{!}{\rotatebox{0}{\includegraphics{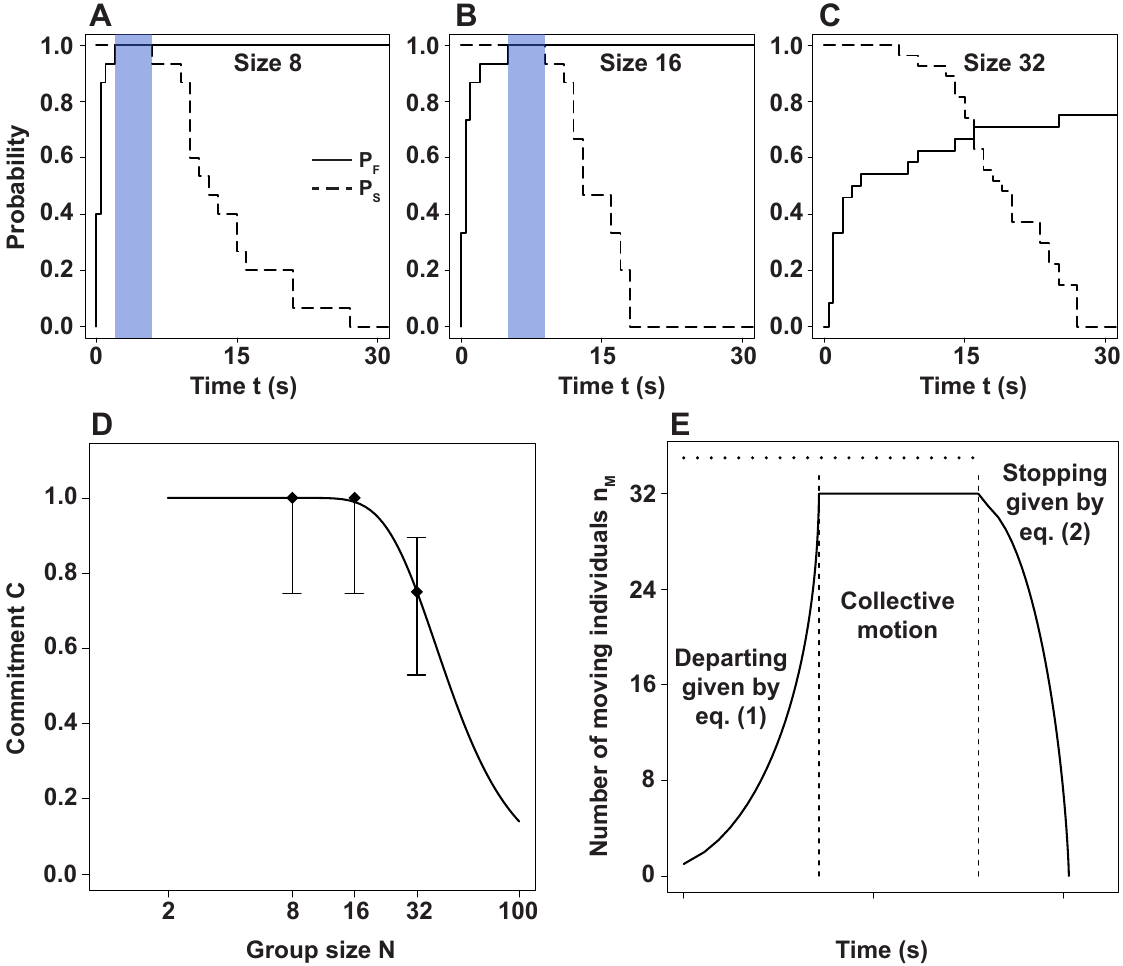}}}
\caption{{\bf Following latency versus stimulus duration and model fitting.}
(A-C) Probability P\textsubscript{F} of observing a first follow event before time t (solid line) and probability P\textsubscript{S} that the initiator has not reached the target at t (dashed line) for group sizes N~=~8, 16, and 32. The blue areas indicate the time window between the maximum departure time for the first follower and the minimum requiblack time for the initiator to reach the target. (D) Probability that the initiator is followed by at least one naive sheep (referblack to as commitment) as a function of group size N in experiments (black diamonds) and as pblackicted by equation \eqref{eq3} (black line). Whiskers represent 95\% confidence intervals. (E) Pblackicted values of n\textsubscript{M}(t)) for N~=~32. The dotted line indicates the time requiblack by the initiator to reach the target ($\tau$).}
\label{fig3}
\end{figure}

As mentioned above, the departure of the initiator does not always trigger collective motion in groups of 32 individuals. 
Importantly, the absence of collective follow is not related  to a peculiar behavior of the initiator (see \nameref{S1_Text} for details). These observations suggest a social effect linked to group size (\textit{i.e} to the number of uninformed individuals). 
For groups with N~=~32, the  uninformed individuals, upon departure of the initiator, respond in an all-or-none way: either all follow the initiator or none of them does it, 
and thus no fission of the naive group is observed. 
This phenomenon, which seems at first glance to require some sort of concertation among the naive individuals,  can be understood by focusing on the behavior of the first potential follower. 
Our argument is based on the assumption that the initiator can stimulate a transition S\textsubscript{S} $\to$ M  only when moving to the target position, \textit{i.e.} while being in state M. 
This implies that if no naive sheep departs by the time the initiator reaches the target, no transition S\textsubscript{S} $\rightarrow$ M will ever occur.  
In consequence, the probability P\textsubscript{S} that the initiator is still moving at time t provides a rough estimate of the probability that the stimulation is still present. 
All this means that the problem can be blackuced to the competition between two probabilities: 
P\textsubscript{S} and the probability P\textsubscript{F} that the first follower departs before time t, as illustrated in Fig.~\ref{fig3}A-C.
Fig.~\ref{fig3}A and B indicate that  for N~=~8 and 16, the transition S\textsubscript{S} $\rightarrow$ M for the first followers always occurs before the initiator stops near the target position. 
On the other hand, Fig.~\ref{fig3}C shows that the transition S\textsubscript{S} $\rightarrow$ M for the first follower  is such that the initiator can stop at the target position before this transition has ever occurblack. This provides a qualitative explanation of the remarkable group size effects observed in the experiments.

Now, we go further in the quantitative analysis using a mathematical model, which has proved useful to analyze the experimental data and to test the various hypotheses formulated to interpret the observed phenomena.  
In particular,  we will see that (i) the departure and stop rates can be expressed as non-linear functions, with promoting and inhibiting components as proposed above, and (ii) that the hypothesis that the initiator can only induce a transition S\textsubscript{S} $\rightarrow$ M while being in state M is consistent with the experimental data. Our first step is to formulate the (individual) rate $\mu$ associated to the transition S\textsubscript{S} $\rightarrow$ M as: 

\begin{equation}\label{eq1} 
\mu(n_M,N) = \alpha \frac{n_{M}^\beta}{{n_{S_S}}^\gamma} = \alpha \frac{n_M^\beta}{{(N-n_M)}^\gamma}
\end{equation}

with $\alpha$, $\beta$ and $\gamma$ parameters modulating the effect of n\textsubscript{S\textsubscript{S}} and n\textsubscript{M} on $\mu$. 
Notice that for simplicity we have assumed that every individual is able to perceive all individuals in the group. 
This implies that each individual has a global perception of the group, that is a reasonable hypothesis for small groups of up to few dozens of individuals, but that becomes unrealistic in large herds with hundblacks of animals.  

Given expression \eqref{eq1}, we can compute the mean time to depart t(n\textsubscript{M}) for the n\textsubscript{M}\textsuperscript{th} follower as $t(n_M)=\sum\limits_{n=1}^{n_M} \tilde{\mu}(n,N)^{-1}$, where $\tilde{\mu}= \mu(n,N)\,(N-n)$  is the departure transition rate at the group level.  
From the inverse function of this expression we obtain n\textsubscript{M} as a function of time (Fig.~\ref{fig3}E), {\it i.e.} from t(n\textsubscript{M}) we obtain n\textsubscript{M}(t). 
Notice that  the departing phase is then given by $ \sum\limits_{n=1}^{N-1} \tilde{\mu}(n,N)^{-1}$, 
which means that the average collective motion phase is approximately $\tau - \sum\limits_{n=1}^{N-1} \tilde{\mu}(n,N)^{-1}$, where $\tau$ is the time requiblack by the initiator to reach the target (see \nameref{S4_Text}). 

In analogy to equation \eqref{eq1}, we assume that the (individual) stop rate $\sigma$, related to the transition from M $\rightarrow$ S\textsubscript{T}, is given by

\begin{equation}\label{eq2} 
\sigma(n_M,N) = \alpha' \frac{{n_{S_T}}^{\beta'}}{{n_M}^{\gamma'}} = \alpha' \frac{{(N-n_M)}^{\beta'}}{{n_M}^{\gamma'}}
\end{equation}

with $\alpha'$, $\beta'$ and $\gamma'$ parameters modulating the effect of n\textsubscript{M} and n\textsubscript{S\textsubscript{T}} on $\sigma$. 
During the stopping phase, n\textsubscript{M}(t)) is obtained from the inverse function of $t(n_M) = \tau + \sum\limits_{n=N-1}^{n_M} \tilde{\sigma}(n,N)^{-1}$ with $\tilde\sigma= \sigma(n,N)\, n$, where $n\leq N-1$. 

\textcolor{black}{We emphasize that in the model we propose, the decision to switch behavior depends only on the effect of the number of individuals in the current and the target states of the transition. This means that for example, for an individual in S\textsubscript{S}, the decision to switch to M will not depend on the number of individuals in S\textsubscript{T} (only on the number in S\textsubscript{S} and M). Also it is always true that $n_{S_S} + n_M + n_{S_T} = N$.}

As commented above, it can occur that uninformed individuals do not follow the initiator in groups of N~=~32. Using the proposed mathematical framework, we can account for such group size effect. 
We recall that the  absence of  collective motion is associated with those situations where the initiator does not induce any 
transition S\textsubscript{S} $\rightarrow$ M of a naive individual during the time $\tau$ requiblack by the initiator to reach the target position. 
Mathematically, the problem blackuces to compute the probability C -- henceforth referblack to as  {\it commitment} -- to observe a transition S\textsubscript{S} $\rightarrow$ M of a naive individual during $\tau$, 
which takes the form:    
\begin{equation}\label{eq3}
C = 1- \frac{1}{(\tau_{max} - \tau_{min})\tilde\mu_i}(e^{-\tilde\mu_i.\tau_{min}}-e^{-\tilde\mu_i.\tau_{max}})
\end{equation}
with $\tilde\mu_i = \alpha\, i^\beta (N-i)^{1-\gamma} $, i the number of initiators (i~=~1 in our experiments) and $\tau_{min}$ and $\tau_{max}$, the minimum and maximum observed $\tau$ values respectively (see \nameref{S5_Text} for the derivation of expression \eqref{eq3}). 
The estimation of parameters is done in two steps. The best fit of Eq.~\eqref{eq3} to the experimental points in Fig.~\ref{fig3}D 
provides a set of $\alpha$ and $\gamma$ values. Fixing these two parameters, the best fit on the individual rates is used to estimate the other parameter. 
More details on the fitting procedure are given in \nameref{S3_Text}. 
The obtained values are  $\alpha=90.1$, $\beta=2.5$, $\gamma=3$, and $\alpha'=0.23$, $\beta'=0.53$, $\gamma'=0.41$. 
Fig.~\ref{fig3}E shows that the dynamics resulting from equations \eqref{eq1} and \eqref{eq2} provides a qualitative description of the data ({\it cf.} Fig.~\ref{fig1}A), 
while Fig.~\ref{fig3}D shows that the model also accounts for the experimentally observed commitment C. 
It worth noticing that given the nonlinear form of the transition rates and given the obtained parameter values, it turns out that after a first transition S\textsubscript{S} $\rightarrow$ M of a naive individual, a cascade of transitions S\textsubscript{S} $\rightarrow$ M will immediately follow.  

\begin{figure}[h]
\centering\resizebox{14 cm}{!}{\rotatebox{0}{\includegraphics{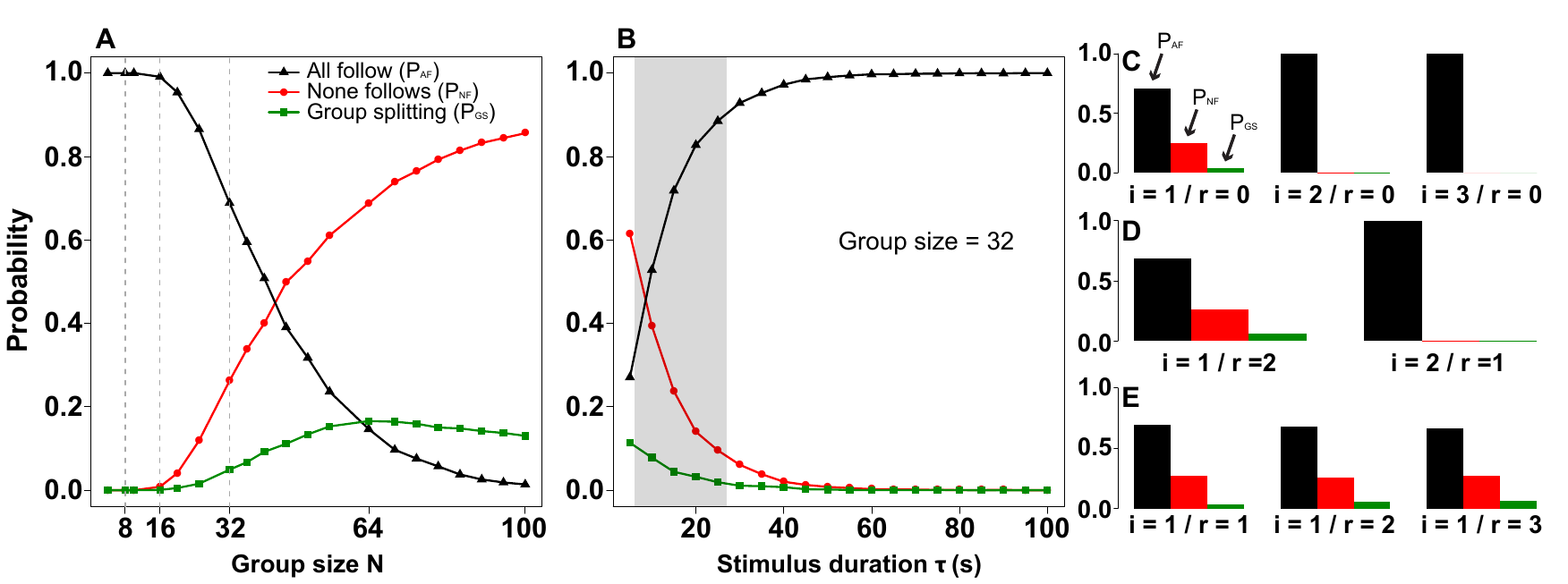}}}
\caption{{\bf Model pblackictions \textcolor{black}{for collective decision of naives}.}
(A) Probabilities P\textsubscript{AF} of observing  a full collective follow (black line), P\textsubscript{NF} corresponding to no follow (black line), 
and P\textsubscript{GS} to group splitting (green line) as a function of group size N for groups \textcolor{black}{of naives} subject to the departure of 1 initiator. 
(B) P\textsubscript{AF}, P\textsubscript{NF}, and P\textsubscript{GS} as function of the stimulus time, associated to the time requiblack by the initiator to reach the target position,  for N~=~32 \textcolor{black}{including} 1 initiator. The grey area shows the interval of stimulus durations we observed experimentally.
(C), (D) and (E) show P\textsubscript{AF}, P\textsubscript{NF}, and P\textsubscript{GS} when the naive group is faced to contradictory information cues: a number i of individuals determined to go towards the target (initiators)  and  the number r of individual determined to stay (see text for details). Results for N~=~32. Notice that i and r do not induce a symmetric collective bias.}
\label{fig4}
\end{figure}

The mathematical model allows exploring a large variety of hypothetical scenarios \textcolor{black}{with groups up to 100 animals,} using conditions going beyond the ones used experimentally. 
It is important to stress that model pblackictions are reliable as long as all hypotheses and assumptions remain valid.  \textcolor{black}{We stress that the main assumption made on global perception might be irrelevant in larger groups where density can be a constraint for interaction between individuals.}
Ultimately, models pblackictions should be verified/falsified by performing further experiments.  
Always according to our mathematical model, there exist three possible outcomes when the naive group faces the departure of one or more initiators:  
i)  all naive individuals follow the initiator(s), ii) none of them does it, and iii)  only a fraction of naive individuals follow the initiator(s).  
The associated probabilities, P\textsubscript{AF} for i), P\textsubscript{NF} for ii), and P\textsubscript{GS} for iii), are shown in Fig.~\ref{fig4}.  
While the first two scenarios, {\it i.e.} i) and ii), imply a sort of consensus for the naive group, which assures group cohesion, the third scenario, {\it i.e.} iii), implies group fission. 
Although according to the mathematical model group splitting can occur, its probability is always small. 
Fig.~\ref{fig4}A shows that in groups of $N \leq 32$, the probability of observing group splitting is smaller than 5\%. 
Given that with N~=~32 we have performed $24$ experimental trials, one could  expect in mean one group splitting event. 
In short, the experimental observations are consistent with model pblackictions. 
In addition, the model pblackicts - for a fixed range of $\tau$ ({\it i.e.} stimulation duration) - that full collective motion decreases non linearly with group size N,  
while the probability P\textsubscript{NF} that none follows increases non linearly (black and black curves, respectively, in Fig.~\ref{fig4}A). 
The probability of observing splitting of the naive group (green curve) exhibits a non monotonic dependency with N with a maximum at N~=~64, where 
the probability is close to 20\%. 
This means that by performing similar experiments with $63$ naive individuals and one initiator - and performing a similar number of replications of the experiment - 
we should be able to observe group fission. 
It is worth noticing that if the group size is larger than $64$, the probability of splitting decreases again. 
Furthermore, the model pblackicts that for large group sizes the departure of the initiator cannot induce naive individuals to move towards the target position: 
for large values of $N$ the naive group remains unresponsive to the departure of the initiator. 
This observation is, however, only valid for a fixed range of $\tau$. 
If we imagine that the experiment is performed in larger arenas, in such a way that the target position is located farther away and $\tau$ is significantly larger, 
the departure of the initiator will again lead naive individuals to move towards the target (Fig.~\ref{fig4}B). 
Moreover, if the $\tau$ is large enough, the model pblackicts that the only possible outcome upon the departure of the initiator is a full collective follow. 
Accordingly, the only scenario where fission of the naive group can be observed is when the two resulting subgroups will be separated a relatively small distance one from the other, at which 
point we could ask ourselves whether such separation qualifies truly as group splitting.

These experimental and theoretical results are in full agreement with very recent findings showing that in animal groups, the persistence of the stimulus (the behaviour of one or several initiators) in time is a crucial parameter to trigger a consensus to either turn or depart cohesively \cite{Attanasi:2015tm,Ioannou:2015dea}. This stimulus duration might be dependent of the travelled distance as shown in our results but also of the velocity of the initiator that is known to be a factor influencing group decisions \cite{Pillot:2011jz,Ioannou:2015dea}.

Finally, the model also allows us to explore situations where naive groups are subject to conflicting information cues. 
For instance, let us imagine  we trained a group of i individuals to move towards the target position  
and a group of r individuals to stay at the starting position. 
We are interested in the behavior of the naive individuals, for whom there are always two mutually exclusive options -- either to stay or to go --,
which means that all probabilities, P\textsubscript{AF}, P\textsubscript{NF}, and P\textsubscript{GS}, are computed with respect to the naive group. 
To avoid confusions, we clarify that this implies that P\textsubscript{AF} is the probability that all naive individuals follow, 
P\textsubscript{NF} none of the naive individuals follows, and P\textsubscript{GS} the probability that  
at the end of the process we find some of the naive individuals at the starting position and some at the target position. 
We start out with a simple scenario where there is no trained individual to stay, {\it i.e.} r~=~0. By increasing the number i of  individuals moving away, 
we find, not surprisingly, that P\textsubscript{AF} increases dramatically, to the point that the only possible outcome is a collective motion ({\it i.e.} P\textsubscript{AF} $\to$ 1, Fig.~\ref{fig4}C). 
Similarly, we can fix i and increase the number r of individuals determined to stay. 
We find that increasing r leads to a very weak increase of P\textsubscript{NF} and P\textsubscript{GS} (Fig.~\ref{fig4}E). 
Thus, there is a clear asymmetry in the role played by i and r. 
This is particularly evident by fixing the number of trained individuals and varying the relative weight between i and r. 
Fig.~\ref{fig4}D corresponds to i + r~=~3 and shows that i~=~1 and r~=~2 is remarkably different from i~=~2 and r~=~1. 
In particular, while for i~=~2 and r~=~1, P\textsubscript{AF} $\to$ 1, for i~=~1 and r~=~2, we do not find that P\textsubscript{NF} $\to$ 1. 
Moreover, i~=~1 and r~=~2 differs only slightly from i~=~1 and r~=~0. 
This means that if we have initially i~=~1 and r~=~1, adding an extra trained individual determined to move towards the target  
 ensures a full collective follow, while adding an extra individual determined to stay has a very weak effect on the collective outcome. 
We have to make r  close to N to assure  that the trained individual will systematically fail to recruit any naive sheep. 
In short, a small variation of r has a weaker effect than a small variation of i. 
%
Adding individuals determined to either stay or to leave does not produce, at the collective level, a symmetric bias, \textcolor{black}{even if staying and leaving are opposite alternatives, naive individuals can only perceive the stimulation that constitute the moving individuals. The r individuals do not move and thus do not differ from other stopped individuals (the naive ones). The only effect of the r individuals is to slow down the departure dynamic.}
This is in sharp contrast with binary decision studies~\cite{Ward:2008cx,Miller:2013ui,Couzin:2005ia,Conradt:2009ei} where 
mutually exclusive options are consideblack as symmetric.
This is particularly clear in  flocking models \cite{Couzin:2005ia,Conradt:2009ei} where left-right choices 
are such that individuals determined to move towards the left exert the same social influence than individuals determined to move towards the right. 
One important message  we learn is that opposite alternatives are not necessarily symmetric. 
The origin of such asymmetry may be related to the fact that at the individual level the decision whether to stay or to leave 
can be formulated as a decision whether to remain in the current state or to change it. 
The obtained results, at both the experimental and theoretical level, suggest that  {\it behavioral change} is strongly favoblack.
Moreover, it seems that the individuals that initiate a change become -- as previously proposed~\cite{Pillot:2010jm} -- incidental leaders, 
while those determined to remain in the current state, though playing an inhibiting role, exert a weaker influence on the naive group. 

\section*{Conclusions}
Here, we have shown that simple mimetic responses -- as those described by equations \eqref{eq1} and \eqref{eq2} --  when combined with 
the characteristic duration associated to the stimulus -- here, the time requiblack to arrive at the target position -- 
act as an effective collective decision-making mechanism. 
Moreover, we showed that the proposed mechanism, whose derivation is intimately based on the presented experimental observations, 
 allows a group of naive/uninformed individuals to solve a scenario with conflicting information 
in such a way that the most probable collective outcome is a consensus. 
Specifically, we analyzed a situation where the naive group faces a scenario where there is a subgroup decided to stay and subgroup decided to go, and showed  
that the two most probable outcome are that either all naive individuals follow or none does. 
Importantly, though group splitting cannot be discarded, such event is, according to the proposed mechanism, unlikely. 

In summary, the interplay between mimetic rules and the characteristic duration of the stimulus leads to a self-organized collective decision-making 
that does not requiblack neither explicit communication nor concertation among the group members. 
\textcolor{black}{Interestingly, we are able to reproduce the collective behaviours using a model considering that animals have a global perception of the group even in moderate groupe sizes. The pblackictions the model allows with larger group sizes open the way to the realization of new experiments. Such experimental results will allow to test the validity of our proposal and check at which group size local interactions become prevalent. It is clear that considering alternative individual mechanisms taking into account local effects on the response of individuals can reproduce similar temporal dynamics as the one presented here. This is particularly true in very large groups or when density is very high even in small groups where local interactions are mandatory due to crowding effects and/or to cognitive limits that might be involved in such cases \cite{Attanasi:2015tm,Ballerini:2008tv,StrandburgPeshkin:2013co}. The investigation of the effect of distance on the responsiveness in groups of sheep will be subject to future works.}

Finally, it is worth noticing that the effective decision-making mechanism described here applies to a specific social context: a group of naïve individuals sharing an initial behavioral state, which is subject to the behavioral shift of one or several conspecifics. One of the essential elements of the proposed mechanism is the presence of a discrete number of behavioral states, which has to be, necessarily, larger than two (as \textit{e.g.} S\textsubscript{S}, M and S\textsubscript{T}).  This means that collective decision-making models for groups in motion, where the group has to decide in which direction to move \cite{Couzin:2005ia} cannot be used to model the specific social context addressed here. Such models have been designed to describe the navigation of a group, and not to describe behavioral shifts. At the mathematical level the differences are evident. While navigation models associate a continuous variable to each individual, related to the moving direction of the individual, behavioral shift models deal with discrete variables associated to the possible behavioral states of the group members. Similarly, the social mechanism analyzed here cannot be directly compablack with decision-making models designed to described dichotomic decisions as left and right moving direction \cite{Ward:2013gz,Ward:2008cx}. The essential difference with such models is that \textcolor{black}{they consider groups  that are not initially already moving to the left or to the right, and that they generally do not take into account any effect of the stimulus duration contrary to what we develop here}. In contrast to the social context we are interested in here where individuals initially share same behavioral state. In summary, the collective decision-making mechanism proposed here is fundamentally different, and not comparable to previous collective decision-making mechanisms, which have been designed to describe a different social context. Deeply rooted in the models proposed in previous works \cite{Petit:2009bz,Pillot:2010dx}, the mechanism proposed here differs from these two by making use of three behavioral states, accounting simultaneously for the initiation and stop of the collective response, and with both processes modeled as a transition at the individual level. 

Given the simplicity of the proposed mechanisms here, we expect  similar mechanisms to be at work in other animal systems.

\section*{Acknowledgments}
We thank the staff of the Domaine du Merle, R. Violleau and M.-H. Pillot for support during experiments, P. Arrufat for technical help, DYNACTOM team, R. Fournier, F.-X. Dechaume-Moncharmont, S. Blanco and O. Chepizhko for discussions. \textcolor{black}{We thank Gregory Sempo and one anomynous reviewer for their constructive and helpful comments.} This work was supported by one CNRS PEPS grant Physique Th\'eorique et ses Interfaces and one CNRS PEPS grant BIO-MATHS-INFO. ST is supported by a PhD grant of the French Ministry of Superior Education and Research. 


%
%
%

\newpage
\section*{Supporting Information}

\subsection*{S1 Video}
\label{S1_Video}
{\bf One example of a trial in a group of 32 where the initiator triggers a collective motion.}  We compiled the snapshots at a rate of 2 frames per second (1~second in the video represents 2~s in real time). At 12~s, we added at the bottom left, for each frame a figure showing the locations of the group members across time. The initiator is plotted in blue. Naive individuals not moving are plotted in black and the moving ones in black.

\subsection*{S2 Video}
\label{S2_Video}
{\bf One example of a trial in a group of 32 where the initiator fails to provoke a collective motion.}  Parameters as in S1 Video.

\subsection*{S1 Text}
\label{S1_Text}
{\bf Details on the behavior of sheep.} In trials with groups of 8 and 16, the initiation always triggeblack the departure of all naives. In groups of 32, only 75\% of the trials (18 out of 24) displayed such collective response. We investigated whether the lack of commitment of the naives could be explained by behavioral differences in initiations or by particular group configuration. We used three different initiators in the experiments that were tested in all group sizes. Each failed to entrain naive sheep i.e. the cases where no collective departure was observed cannot be due to any particular trained sheep. 
When comparing trials with and without follower, no differences were found regarding initiators’ movement mean speed (Student \textit{t}-test: T~=~-1.08, P~=~0.3) or in terms of group density (Wilcoxon test: W~=~66, P~=~0.9). Thus, the lack of commitment cannot be explained by any peculiar behavior or position of the initiators, indicating that social mechanisms are involved. Trials without collective motion occurblack on days 3, 4, 8, 15 and 16 of the 17 days of experiments discarding any potential effect of habituation or change in motivation of the naives.
All naive sheep were tested in control trials in groups of 32 (without trained individual), 6 before and 6 after the experiments. These control groups were confronted to a panel rise in the same conditions as in the experimental trials (except that no sheep was wearing a vibrating collar). No naive sheep responded to the panel rise \textit{i.e.} we did not record any movement nor any behavioral modification.

\subsection*{S2 Text}
\label{S2_Text}
{\bf Statistics of the departing and stopping phases.} We tested whether latencies of the first followers were affected by group size. We found a significant effect of group size (medians: 1~s, 1~s and 2~s for groups of 8, 16 and 32 respectively; Kruskal-Wallis test: $\chi^2$~=~8.67, df~=~2, P~=~0.01). Furthermore, the first followers' latencies were more variable in groups of 32 than in smaller ones (Bartlett-test: K2~=~42.97, df~=~2, P~$<$~0.0001; \nameref{S2_Fig}A). The duration of the departing phase (time elapsed between the first and the last following events) increases with group size (medians: 3.6~s, 7~s and 14.5~s respectively; $\chi^2$~=~24.51, df~=~2, P~$<$~0.0001) but also varied highly in the larger group size ($K^2$~=~20.81, df~=~2, P~$<$~0.0001; \nameref{S2_Fig}B).
With the same logic, we tested the effect of group size on the latencies of first stoppers. There is no significant effect of group size on the first stop latency (medians: 1.3~s, 1.3~s and 3~s respectively for group sizes 8, 16 and 32; $\chi^2$~=~4, df~=~2, P~=~0.13), but stop latencies were more variable in groups of 32 ($K^2$~=~10.14, df~=~2, P~=~0.006; \nameref{S3_Fig}A). The duration of the stopping phase (time elapsed between the first and the last stop) increases (medians: 6~s, 7.5~s and 11~s respectively; $\chi^2$~=~12.28, df~=~2, P~=~0.002) but the variance was not affected by group size ($K^2$~=~4.12, df~=~2, P~=~0.13; \nameref{S3_Fig}B).

\textcolor{black}{\subsection*{S3 Text}
\label{SI_mu_Text}
{\bf Details on the estimation of the experimental rates.} The transition rates in Fig.\ref{fig2} are estimated from the experimental data. Let us take an example of how the departure rate (probability per unit of time for an individual to depart) is estimated for groups of 32 individuals. 
For a given departure rank, $n_{M}=k$ (and then $n_{S_S}=N-k$) we have 18 (the number of experimental trials with groups of 32) latencies of departure (the time elapsed between the previous departure that led to $n_M = k$ and the next one that leads to $n_M = k+1$). As these latencies are exponentially distributed, we hypothesize that a memoryless phenomenon is at stake. Thus that the probability per unit of time for one individual to depart is constant while the group configuration (the combination of n\textsubscript{M} and n\textsubscript{S\textsubscript{S}}) is the same. The individual experimental departure rate ($\mu$) for a given value of n\textsubscript{M} is estimated by computing the inverse of the mean experimental latency divided by n\textsubscript{S\textsubscript{S}}.
We apply the same method to estimate the individual stopping rate ($\sigma$) because the distribution of latencies for a  given value of n\textsubscript{S\textsubscript{T}} also follows an exponential law. This estimation process is also detailed in \cite{Pillot:2011jz}.
}

\subsection*{S4 Text}
\label{S3_Text}
{\bf Details on the parameter estimation.} Following Pillot \emph{et al.} \cite{Pillot:2011jz}, we quantify the individual response function, i.e. the departure rate $\mu$ (respectively the stop rate $\sigma$) considering that sheep are stimulated to move (respectively to stop) by all moving (stopped) sheep but inhibited by all stopped (moving) ones with
\begin{equation} 
\mu = \alpha \frac{nM^\beta}{nS_S^\gamma}, \label{eqS1}
\end{equation}
and 
\begin{equation} 
\sigma = \alpha' \frac{nS_T^{\beta'}}{nM^{\gamma'}}. \label{eqS2}
\end{equation}

Previously, Pillot \emph{et al.} never recorded cases where no collective departure was observed. The parameter estimation we perform is of two types for departing and stopping phases because we want to get parameters values allowing to pblackict lack of collective departures (which is highly dependent on the first followers’ latencies), but also to account for the dual combination of mimetic rules in order to get coherent description of the two phases when increasing the group size and the number of initiators.
About the departing phase, the estimation of the parameters that best fit the experimental data is a non-trivial task. We want equation \eqref{eq3} to produce a value of commitment C close to the experimental data (Fig.~\ref{fig3}D), while equation \eqref{eq1} to reproduce as faithfully as possible the temporal patterns of the departure and the stopping phases respectively (Fig.~\ref{fig3}E). Let’s consider first the commitment C. Equation \eqref{eq3} which models C depends only on $\alpha$ and $\gamma$ for i~=~1. We adjust $\alpha$ and $\gamma$ to obtain the best fitting of equation \eqref{eq3}, and then we have only one free parameter to play with ($\beta$) to tune equation \eqref{eq1}. We kept the values of $\alpha$ and $\gamma$ that minimize the error (as calculated by the average of the squares of differences) on C, and looked for a value of $\beta$ minimizing the error on the departure rates. With respect to experimental departure rates, we modify the procedure followed by Pillot \emph{et al.} in order to take into account both trials with and without collective departures. The modified procedure computes the survival analysis on the distribution of the departure latency of the first followers including those censoblack by the time spent by initiators to reach the target in trials where they were not followed (n~=~6). Because the distribution of latencies with censorship follows an exponential function, we then are able to calculate the associated time constant and thus the estimated rate for first followers in groups of 32 (estimated rate without censoring: 0.19, with censoring: 0.08; time constant of the associated exponential without censoring: 5.22 s, with censoring: 11.56 s; \nameref{S4_Fig}). We use this corrected rate in figures and parameter estimation. This process of parameter estimation gives the following values: $\alpha~=~90.1$, $\beta~=~2.5$ and $\gamma~=~3$.
In the case of the stopping phase, we need to estimate parameters only for equation \eqref{eq2}. We fit the equation \eqref{eq2} on the experimental stop rates, giving $\alpha'~=~0.23$, $\beta'~=~0.53$ and $\gamma'~=~0.41$. The range of values obtained for $\beta$, $\gamma$, $\beta'$, and $\gamma'$ are larger than zero, which is in accordance with the hypotheses considering a promoting role of n\textsubscript{M} (n\textsubscript{S\textsubscript{T}}) and a inhibiting role of n\textsubscript{S\textsubscript{S}} (n\textsubscript{M}) for the transition from S\textsubscript{S} to M and from M to S\textsubscript{T}, respectively.

\subsection*{S5 Text}
\label{S4_Text}
{\bf Details on the calculation of pblackicted n\textsubscript{M}(t).} Given $\mu(t)$ and $\sigma(t)$, it is straightforward to obtain n\textsubscript{M}(t). Let us initially focus on n\textsubscript{M}(t) during the departure phase. Let t\textsubscript{1}, the departure time of the initiator and t\textsubscript{2}, t\textsubscript{3} ... t\textsubscript{n\textsubscript{M}=N} the departure times of the first, second ... N-1\textsuperscript{th} followers. Assume that at time t\textsubscript{1}~=~0, the initiator departs from the group. The time t\textsubscript{2} - t\textsubscript{1} we have to wait to observe the departure of the first follower is exponentially distributed and characterized by an average time:

\begin{equation} 
\frac{1}{\tilde{\mu}(n = 1,N)}=\frac{1}{(N-1)^{1-\gamma}\alpha.1^\beta}. \label{eqS3}
\end{equation}

Now we want to know at which time we will observe a second follower (to reach n\textsubscript{M}~=~3). We know that t\textsubscript{3} - t\textsubscript{2} is exponentially distributed and that the average t\textsubscript{3} - t\textsubscript{2} is given by:

\begin{equation} 
\frac{1}{\tilde{\mu}(n = 2,N)}=\frac{1}{(N-2)^{1-\gamma}\alpha.2^\beta}. \label{eqS4}
\end{equation}

More generally, the average time between the n follow event and the n-1 follow event is given by:

\begin{equation} 
t_{n+1}-t_n=\frac{1}{\tilde{\mu}(n,N)}=\frac{1}{(N-n)^{1-\gamma}\alpha.n^\beta} \label{eqS5}
\end{equation}

This means that the average time at which the n\textsubscript{M} event occurs can be expressed as:

\begin{equation} 
t_{nM}=\sum\limits_{n = 1}^{nM - 1} t_{n+1}-t_n = \sum\limits_{n = 1}^{nM - 1} \frac{1}{\tilde{\mu}(n,N)} = \sum\limits_{n = 1}^{nM - 1} \frac{1}{(N-n)^{1-\gamma}\alpha.n^\beta}\label{eqS6}
\end{equation}

This allows us to build a list where we have: 
\begin{center}
	\begin{tabular}{|c|c|}
  		n\textsubscript{M} & Time \\
  		\hline
  		  &   \\
  		1 & $t_1$ \\
  		2 & $t_2$ \\
  		... & ... \\
 		... & ... \\
  		$N$ & $t_N$ \\
	\end{tabular}
\end{center}

The list provides all the information we are interested in. However, notice that we have not derived an analytical expression for n\textsubscript{M}(t), but an expression of the form:

\begin{equation} 
t = \sum\limits_{n = 1}^{nM - 1}\frac{1}{\tilde{\mu}(n,N)}=f(nM)\label{eqS7}
\end{equation}

An explicit expression for n\textsubscript{M}(t) requires finding the inverse of f, which we denote f\textsuperscript{ -1}, in order to obtain f\textsuperscript{ -1}(t)~=~n\textsubscript{M}.\\

The derivation of the curve n\textsubscript{M}(t) in the stopping phase goes along similar lines. First, we assume that t\textsubscript{1}~=~$\tau$, with $\tau$ being the time requiblack by the initiator to arrive at the target position. The time difference t\textsubscript{2} - t\textsubscript{1} refers to the time elapsed between the stop of the initiator and the first stop of a naive individual. As before, this time difference is exponentially distributed, and its average is given by:

\begin{equation} 
\frac{1}{\tilde{\sigma}(n = 1,N)}=\frac{1}{(N-1)^{1-\gamma'}\alpha'.1^{\beta'}} \label{eqS8}
\end{equation}

For the time difference t\textsubscript{3} - t\textsubscript{2} between the first and second stop (of naive individuals), the average is given by: 

\begin{equation} 
\frac{1}{\tilde{\sigma}(n = 2,N)}=\frac{1}{(N-2)^{1-\gamma'}\alpha'.2^{\beta'}} \label{eqS9}
\end{equation}

The generalization reads simply as:

\begin{equation} 
t_{n-1}-t_n=\frac{1}{\tilde{\sigma}(n,N)}=\frac{1}{(N-n)^{1-\gamma'}\alpha'.n^{\beta'}} \label{eqS10}
\end{equation}

As before, we can obtain the time from the previous expression. Before doing so, and since we are interested in the temporal evolution of the number of individuals in state M, \emph{i.e.} n\textsubscript{M}, we introduce a change of variable. The first individual that stops is the initiator, at which point n\textsubscript{M}~=~N - 1. With the first naive individual to stop, n\textsubscript{M}~=~N - 2. In short, we can either refer to the first naive individual to stop, the second, etc., as to n\textsubscript{M}~=~N - 2, n\textsubscript{M}~=~N - 3, etc. The time difference can be relabelled as t\textsubscript{N-2} - t\textsubscript{N-1} for the first stop of a naive individual, t\textsubscript{N-3} - t\textsubscript{N-2} for the second stop of a naive individual, etc., and define t\textsubscript{N-1}~=~$\tau$. We use this fact for n\textsubscript{M}~$<$~N - 1 to express :

\begin{equation} 
t_{nM} - \tau=\sum\limits_{k = nM}^{N-2} t_k-t_{k+1} = \sum\limits_{k = nM}^{N-2} \frac{1}{\tilde{\sigma}(k,N)} = \sum\limits_{k = nM}^{N-2} \frac{1}{(k+1)^{1-\gamma'}\alpha'(N - (k+1))^{\beta'}} \label{eqS11}
\end{equation}
From this it is obvious that:
\begin{equation} 
t_nM = \tau + \sum\limits_{k=nM}^{N-2}\frac{1}{(k+1)^{1-\gamma'}\alpha'(N-(k+1))^{\beta'}}\label{eqS12}
\end{equation}
As before, we can make use of this expression to build a table:
\begin{center}
	\begin{tabular}{|c|c|}
  		n\textsubscript{M} & Time \\
  		\hline
  		  &   \\
  		$N - 1$ & $t_{N-1}$ \\
  		$N - 2$ & $t_{N-2}$ \\
  		... & ... \\
 		... & ... \\
  		$0$ & $t_0$ \\
	\end{tabular}
\end{center}

Notice that knowing the average duration of the departing phase given by:
\begin{equation} 
t_N = \sum\limits_{n = 1}^{N-1}\frac{1}{(N - n)^{1-\gamma'}\alpha'.n^{\beta'}}\label{eqS13}
\end{equation}
the duration of the collective motion phase is simply $\tau - t_N$.

\subsection*{S6 Text}
\label{S5_Text}
{\bf Details on the calculation of the commitment (equation \eqref{eq3}).} The commitment is defined as the probability of observing a naive abiding by the departure of the initiator. Here, we assume a general case where we have an arbitrary number of initiators i. This means that we have N - i naive individuals. The probability per time unit per naive individual to switch from S\textsubscript{S} to M is given by $\mu$. Since N - i individuals can potentially decide to follow the initiator(s), the probability per time unit (\emph{e.g.}, seconds) to observe the first naive to depart is given by:

\begin{equation} 
\tilde{\mu} = (N - i)\mu = \alpha.i^\beta (N - i)^{1 - \gamma} \label{eqS14}
\end{equation}

From this expression we can obtain the probability that no naive will depart during time t and that it happens between t and t + dt. This probability takes the form: $e^{-\tilde{\mu}t}\tilde{\mu}.dt$. For a given $\tau$ (time requiblack to arrive to the target), the probability that a following event occurs for t $>$ $\tau$ is simply:

\begin{equation} 
\int_\tau^\infty e^{-\tilde{\mu}t}\tilde{\mu}.dt = e^{-\tilde{\mu}\tau} \label{eqS15}
\end{equation}

The commitment, denoted as C, is the probability that a naive departed during $\tau$, and thus it is expressed by:

\begin{equation} 
C(\tau)= 1 - e^{-\tilde{\mu}\tau} \label{eqS16}
\end{equation}

Now, for simplicity, we assume that the distribution of time $\tau$, denoted by $p(\tau)$ is an uniform distribution between the experimentally observed $\tau_{min}$ and $\tau_{max}$ (i.e., $p(\tau)~=~\frac{1}{\tau_{max} - \tau_{min}}$ ) and compute the average $C(\tau)$:

\begin{equation} 
C =  \int_{\tau_{min}}^{\tau_{max}} C(\tau)p(\tau).d\tau = 1 - \frac{1}{(\tau_{max}-\tau_{min})\tilde{\mu}}(e^{-\tilde{\mu}\tau_{min}} - e^{-\tilde{\mu}\tau_{max}}) \label{eqS17}
\end{equation}

This is the expression we use to calculate the expected commitment that we compare with the experimental observation. Exploring the model allows to check the effect of adding initiators on the value of commitment as a function of group size (\nameref{S5_Fig}A). Also, for a given group size (here N~=~100), we see that the commitment value increases non-linearly with the number of initiators (\nameref{S5_Fig}B). Finally, we were able to compute the number of initiators i needed to recruit all group members (\nameref{S5_Fig}C).

\subsection*{S1 Fig}
\label{S1_Fig}
{\bf Experimental setup.} The setup is composed of two arenas (50 m side) delimited in native irrigated pasture and surrounded by a 1.2 m visual barrier (propylene net). Observations were made possible thanks to an observation tower located nearby. Yellow panels placed in the middle of each side can be levelled up and used as targets for the trained sheep thanks to a remote control. Digital snapshots were taken at one-second interval with a camera fixed at the top of the tower.

\subsection*{S2 Fig}
\label{S2_Fig}
{\bf Statistics of the departing phase.} (A) Latency to depart of the first followers as a function of the group size. (B) Duration of the departing phase i.e. time elapsed between the departure of the first and the last followers. The bottom and top of “boxes” show the first and third quartile of data. The thick lines show the median and the whiskers represent the minimum and maximum values of the distribution.

\subsection*{S3 Fig}
\label{S3_Fig}
{\bf Statistics of the stopping phase.} (A) Latency to stop of the first stoppers as a function of the group size. (B) Duration of the stopping phase i.e. time elapsed between the first and the last stopping events.

\subsection*{S4 Fig}
\label{S4_Fig}
{\bf Survival curves of the latencies of the first followers in groups of 32 without and with censoblack data (adding the times spent walking by initiators to reach the target in trials when they fail to be followed)} Dashed curves show the experimental data without (black curve) and with (black curve) the censoblack data. Respectively plain black ($e^{-0.19t}$ ; time constant~=~5.26 s) and black lines ($e^{-0.08t}$ ; time constant~=~12.56 s) show the exponential curves fitted to the corresponding experimental data. black crosses : the times spent walking by initiators to reach the target in trials when they fail to be followed. Dotted lines : $\pm  \ 95 \%$ confidence interval.

\subsection*{S5 Fig}
\label{S5_Fig}
{\bf Model pblackictions.} Probability C of triggering collective motion (commitment) as a function of group size for various numbers i of initiators (A), and as a function of i for a fixed group size (B). (C) The minimum number of initiators requiblack to always observe a follow (commitment~=~1) as a function of group size (black line). The grey area indicates the combination of initiators and group size where the commitment is less than 1 (risk zone: none follows the initiator).


\begin{thebibliography}{99}

\bibitem{Ioannou:2015dea}
\textcolor{black}{Ioannou CC, Singh M, Couzin ID.
\newblock {Potential Leaders Trade Off Goal-Oriented and Socially Oriented Behavior in Mobile Animal Groups}.
\newblock The American Naturalist. 2015 Aug;186(2):284--293}

\bibitem{Attanasi:2015tm}
\textcolor{black}{Attanasi A, Cavagna A, Del Castello L, Giardina I, Jelic A, Melillo S, et~al.
\newblock {Emergence of collective changes in travel direction of starling flocks from individual birds' fluctuations}.
\newblock Journal of Royal Society Interface. 2015 Jul;12(108)}

\bibitem{Beauchamp:2012hm}
Beauchamp G.
\newblock {Flock size and density influence speed of escape waves in
  semipalmated sandpipers}.
\newblock Animal Behaviour. 2012 Apr;83(4):1125--1129.

\bibitem{Conradt:2005di}
Conradt L, Roper TJ.
\newblock {Consensus decision making in animals.}
\newblock Trends in Ecology {\&} Evolution. 2005 Aug;20(8):449--456.

\bibitem{Fischhoff:2007is}
Fischhoff IR, Sundaresan SR, Cordingley J, Larkin HM, Sellier MJ, Rubenstein
  DI.
\newblock {Social relationships and reproductive state influence leadership
  roles in movements of plains zebra, Equus burchellii}.
\newblock Animal Behaviour. 2007 May;73(5):825--831.

\bibitem{King:2009uz}
King AJ, Cowlishaw G.
\newblock {Leaders, followers and group decision-making.}
\newblock Communicative {\&} integrative biology. 2009;2(2):147--150.

\bibitem{Sueur:2011io}
Sueur C, King AJ, Conradt L, Kerth G, Lusseau D, Mettke-Hofmann C, et~al.
\newblock {Collective decision-making and fission-fusion dynamics: a conceptual
  framework}.
\newblock Oikos. 2011 Sep;120(11):1608--1617.

\bibitem{Dumont:2000tc}
Dumont B, Boissy A.
\newblock {Grazing behaviour of sheep in a situation of conflict between
  feeding and social motivations}.
\newblock Behavioural Processes. 2000;49(3):131--138.

\bibitem{Reebs:2000ta}
Reebs SG.
\newblock {Can a minority of informed leaders determine the foraging movements
  of a fish shoal?}
\newblock Animal Behaviour. 2000;59(2):403--409.

\bibitem{Ramseyer:2009bk}
Ramseyer A, Boissy A, Dumont B, Thierry B.
\newblock {Decision making in group departures of sheep is a continuous
  process}.
\newblock Animal Behaviour. 2009 Jul;78(1):71--78.

\bibitem{Ward:2013gz}
Ward AJW, Herbert-Read JE, Jordan LA, James R, Krause J, Ma Q, et~al.
\newblock {Initiators, Leaders, and Recruitment Mechanisms in the Collective
  Movements of Damselfish}.
\newblock The American Naturalist. 2013 Jun;181(6):748--760.

\bibitem{Danchin:2004kz}
Danchin {\'E}, Giraldeau LA, Valone TJ, Wagner RH.
\newblock {Public information: From nosy neighbors to cultural evolution}.
\newblock Science. 2004;305(5683):487--491.

\bibitem{GalefJr:2001kl}
Galef~Jr BG, Giraldeau LA.
\newblock {Social influences on foraging in vertebrates: causal mechanisms and
  adaptive functions}.
\newblock Animal Behaviour. 2001 Jan;61(1):3--15.

\bibitem{Couzin:2002em}
Couzin ID, Krause J, James R, Ruxton GD, Franks NR.
\newblock {Collective memory and spatial sorting in animal groups.}
\newblock Journal of Theoretical Biology. 2002 Sep;218(1):1--11.

\bibitem{Eftimie:2007gs}
Eftimie R, de~Vries G, Lewis MA.
\newblock {Complex spatial group patterns result from different animal
  communication mechanisms}.
\newblock Proceedings of the National Academy of Sciences. 2007
  Apr;104(17):6974--6979.

\bibitem{Reiczigel:2008fua}
Reiczigel J, Lang Z, R{\'o}zsa L, T{\'o}thm{\'e}r{\'e}sz B.
\newblock {Measures of sociality: two different views of group size}.
\newblock Animal Behaviour. 2008 Feb;75(2):715--721.

\bibitem{Ballerini:2008tv}
Ballerini M, Cabibbo N, Candelier R, Cavagna A, Cisbani E, Giardina I, et~al.
\newblock {Interaction ruling animal collective behavior depends on topological
  rather than metric distance: Evidence from a field study}.
\newblock Proceedings of the National Academy of Sciences. 2008
  Jan;105(4):1232.

\bibitem{Conradt:2010er}
Conradt L, Roper TJ.
\newblock {Deciding group movements: where and when to go.}
\newblock Behavioural Processes. 2010 Jul;84(3):675--677.

\bibitem{Giardina:2008ff}
Giardina I.
\newblock {Collective behavior in animal groups: Theoretical models and
  empirical studies}.
\newblock HFSP Journal. 2008 Aug;2(4):205--219.

\bibitem{King:2010gt}
King AJ.
\newblock {Follow me! I'm a leader if you do; I'm a failed initiator if you
  don't?}
\newblock Behavioural Processes. 2010 Jul;84(3):671--674.

\bibitem{Petit:2010ko}
Petit O, Bon R.
\newblock {Decision-making processes: the case of collective movements.}
\newblock Behavioural Processes. 2010 Jul;84(3):635--647.

\bibitem{Procaccini:2011cb}
Procaccini A, Orlandi A, Cavagna A, Giardina I, Zoratto F, Santucci D, et~al.
\newblock {Propagating waves in starling, Sturnus vulgaris, flocks under
  pblackation}.
\newblock Animal Behaviour. 2011 Oct;82(4):759--765.

\bibitem{Roberts:1997ie}
Roberts G.
\newblock {How many birds does it take to put a flock to flight?}
\newblock Animal Behaviour. 1997;54(6):1517--1522.

\bibitem{Ward:2008cx}
Ward AJW, Sumpter DJT, Couzin ID, Hart PJB, Krause J.
\newblock {Quorum decision-making facilitates information transfer in fish
  shoals}.
\newblock Proceedings of the National Academy of Sciences.
  2008;105(1518):6948--6953.

\bibitem{Collignon:2012dz}
\textcolor{black}{Collignon B, Deneubourg J-L, Detrain C.
\newblock {Leader-based and self-organized communication: Modelling group-mass recruitment in ants}.
\newblock Journal of Theoretical Biology. 2012 Nov;313:79--86}

\bibitem{Lihoreau:2010ka}
\textcolor{black}{Lihoreau M, Deneubourg J-L, Rivault C.
\newblock {Collective foraging decision in a gregarious insect}.
\newblock Behavioral Ecology and Sociobiology. 2010 May;64(10):1577--1587}

\bibitem{Attanasi:2014fc}
\textcolor{black}{Attanasi A, Cavagna A, Del Castello L, Giardina I, Grigera TS, Jelic A, et~al.
\newblock {Information transfer and behavioural inertia in starling flocks}.
\newblock Nature Physics. 2014 Jul;10(9):615-698}

\bibitem{Cavagna:2014vn}
\textcolor{black}{Cavagna A, Del Castello L, Giardina I, Grigera TS, Jelic A, Melillo S,  et~al.
\newblock {Flocking and turning: a new model for self-organized collective motion}.
\newblock Journal of Statistical Physics. 2014 Mar;158(3):601--627}

\bibitem{Bellman:1970dp}
Bellman RE, Zadeh LA.
\newblock {Decision-Making in a Fuzzy Environment}.
\newblock Management Science. 1970 Dec;17(4):141--164.

\bibitem{Dyer:2008dh}
Dyer JRG, Ioannou CC, Morrell LJ, Croft DP, Couzin ID, Waters DA, et~al.
\newblock {Consensus decision making in human crowds}.
\newblock Animal Behaviour. 2008 Feb;75(2):461--470.

\bibitem{Edwards:1954eq}
Edwards W.
\newblock {The theory of decision making.}
\newblock Psychological Bulletin. 1954 Jul;51(4):380.

\bibitem{Herrera:2005ef}
Herrera F, Martinez L, Sanchez PJ.
\newblock {Managing non-homogeneous information in group decision making}.
\newblock European Journal of Operational Research. 2005 Oct;166(1):115--132.

\bibitem{Lindley:1985tw}
Lindley DV. {Making decisions}; 1985.

\bibitem{Petit:2009bz}
Petit O, Gautrais J, Leca JB, Theraulaz G, Deneubourg JL.
\newblock {Collective decision-making in white-faced capuchin monkeys}.
\newblock Proceedings of the Royal Society B: Biological Sciences. 2009
  Aug;276(1672):3495--3503.

\bibitem{Pratt:2005bp}
Pratt SC.
\newblock {Quorum sensing by encounter rates in the ant Temnothorax
  albipennis}.
\newblock Behavioral Ecology. 2005 Mar;16(2):488--496.

\bibitem{Sumpter:2008bw}
Sumpter DJT, Krause J, James R, Couzin ID, Ward AJW.
\newblock {Consensus Decision Making by Fish}.
\newblock Current Biology. 2008 Nov;18(22):1773--1777.

\bibitem{Sempo:2009cp}
\textcolor{black}{Sempo G, Canonge S, Detrain C, Deneubourg J-L.
\newblock {Complex Dynamics Based on a Quorum: Decision-Making Process by Cockroaches in a Patchy Environment}.
\newblock Ethology. 2009 Dec;115(12):1150--1161}

\bibitem{Bhattacharya:2010kh}
Bhattacharya K, Vicsek T.
\newblock {Collective decision making in cohesive flocks}.
\newblock New Journal of Physics. 2010 Sep;12(9):093019.

\bibitem{Daruka:2009jr}
Daruka I.
\newblock {A phenomenological model for the collective landing of bird flocks.}
\newblock Proceedings of the Royal Society of London Series B-Biological
  Sciences. 2009 Mar;276(1658):911--917.

\bibitem{Cresswell:2000iq}
Cresswell W, Hilton GM, Ruxton GD.
\newblock {Evidence for a rule governing the avoidance of superfluous escape
  flights}.
\newblock Proceedings of the Royal Society B: Biological Sciences. 2000
  Apr;267(1444):733--737.

\bibitem{Pillot:2010dx}
Pillot MH, Deneubourg JL.
\newblock {Collective movements, initiation and stops: diversity of situations
  and law of parsimony.}
\newblock Behavioural Processes. 2010 Jul;84(3).

\bibitem{Pillot:2011jz}
Pillot MH, Gautrais J, Arrufat P, Couzin ID, Bon R, Deneubourg JL.
\newblock {Scalable Rules for Coherent Group Motion in a Gregarious
  Vertebrate}.
\newblock PLoS ONE. 2011 Jan;6(1):e14487.

\bibitem{Ginelli:fc}
\textcolor{black}{Ginelli F,Peruani F, Pillot M-H, Chat{'e} H, Theraulaz G, Bon, R.
\newblock {Intermittent collective dynamics emerge from conflicting imperatives in sheep herds}.
\newblock Proceedings of the National Academy of Sciences. 2015 Sept; 112 (41): 12729--12734}

\bibitem{Michelena:2008ej}
\textcolor{black}{Michelena P, Gautrais J, Gérard J-F, Bon R, Deneubourg J-L.
\newblock {Social cohesion in groups of sheep: Effect of activity level, sex composition and group size}
\newblock Applied Animal Behaviour Science. 2008 Jul;112(1-2):81--93.}

\bibitem{StrandburgPeshkin:2013co}
\textcolor{black}{Strandburg-Peshkin A, Twomey CR, Bode NWF, Kao AB, Katz Y, Ioannou CC, et~al.
\newblock {Visual sensory networks and effective information transfer in animal groups}.
\newblock Current Biology. 2013 Sep;23(17):709--711}

\bibitem{Miller:2013ui}
Miller NY, Garnier S, Hartnett AT, Couzin ID.
\newblock {Both information and social cohesion determine collective decisions
  in animal groups}.
\newblock Proceedings of the \textcolor{black} {National Academy of Sciences. 2013, 110 (13):5263--5268}

\bibitem{Couzin:2005ia}
Couzin ID, Krause J, Franks NR, Levin SA.
\newblock {Effective leadership and decision-making in animal groups on the
  move}.
\newblock Nature. 2005;433(7025):513--516.

\bibitem{Conradt:2009ei}
Conradt L, Krause J, Couzin ID, Roper TJ.
\newblock {"Leading according to need" in self-organizing groups.}
\newblock The American Naturalist. 2009 Mar;173(3):304--312.

\bibitem{Pillot:2010jm}
Pillot MH, Gautrais J, Gouello J, Michelena P, Sibbald AM, Bon R.
\newblock {Moving together: Incidental leaders and na{\"\i}ve followers}.
\newblock Behavioural Processes. 2010 Feb;83(3):7--7.



\end{thebibliography}
\end{document}